\documentstyle[amssymb,aps,graphics]{revtex}

\def\be{\begin{equation}}
\def\ee{\end{equation}}
\def\bea{\begin{eqnarray}}
\def\eea{\end{eqnarray}}

\def\ignore#1{ }
\def\bbox#1{ }
\def\bbox#1{  \fbox{\em  #1 } }

\begin{document}
\draft
\title{On the topology of adiabatic passage}
\author{L. P. Yatsenko \thanks{{\it Permanent address: }Institute of Physics,
National Academy of Sciences of Ukraine, prospekt Nauky, 46, Kiev-22, 03650,
Ukraine}, S. Gu\'erin and H. R. Jauslin,}
\address{Laboratoire de Physique de l'Universit\'e de Bourgogne, CNRS,\\
BP 47870, 21078 Dijon, France}
\date{14 dec 2000}
\maketitle

\begin{abstract}
We examine the topology of eigenenergy surfaces characterizing the
population transfer processes based on adiabatic passage. We show that this
topology is the essential feature for the analysis of the population
transfers and the prediction of its final result. We reinterpret diverse
known processes, such as stimulated Raman adiabatic passage (STIRAP),
frequency-chirped adiabatic passage and Stark-chirped rapid adiabatic
passage (SCRAP). Moreover, using this picture, we display new related
possibilities of transfer. In particular, we show that we can selectively
control the level which will be populated in STIRAP process in $\Lambda$ or $%
V $ systems by the choice of the peak amplitudes or the pulse sequence.
\end{abstract}


Email: yatsenko@physik.uni-kl.de, guerin@jupiter.u-bourgogne.fr

\section{Introduction}

Adiabatic passage is now a well established tool to achieve complete
population transfer between discrete quantum states of atoms and molecules.
The main advantage of the processes based on adiabatic passage is their
relative robustness with respect to variation of field parameters. The
adiabatic passage is achieved with adapted adiabatic variations of at least
two {\it effective }parameters of the total laser field. They can be e.g.
the amplitude and the detuning (chirping) or e.g. the amplitudes of two
delayed pulses [stimulated Raman adiabatic passage (STIRAP), see \cite
{STIRAP} for a review]. A chirp can be induced either by direct sweeping of
the frequency of one laser pulse \cite
{Loy,Brewer,Allen,Breuer_PLA,Chelkowski,Guerin_PRA97,Chelkowski_JRS}, or as
proposed very recently by a Stark shift of the transition due to an
additional laser field [process named Stark chirped rapid adiabatic passage
(SCRAP)] \cite{Yatsenko_PRA99,Rickes}. The use of adiabatic passage for
multiple photon absorption and emission processes accompanying momentum
exchanges between the atom and the laser fields have also been recently
investigated \cite{Unanyan_EPJD00,Romanenko_JETP2000,Guerin_PR00}.

In this paper, we show that all of these adiabatic passage processes can be
understood by an analysis of the topology of the surfaces of eigenenergies
as functions of the field parameters. This tool allows us to derive new ways
of controlling population transfer, related to STIRAP.

The tools of analysis follow Ref. \cite{Guerin_PR00}. From an effective
Hamiltonian, which can be constructed by quasi-resonant approximations
combined with adiabatic eliminations from the complete Hamiltonian of the
considered process, we determine and analyze the topology of the energy
surfaces, which display conical intersections and avoided crossings due to
resonances. In the general cases, these resonances are induced by the fields
(dynamical resonances) \cite{Guerin_PR00,Jauslin_PhysicA}. The adiabatic
dynamics of the process is determined by the topology of these energy
surfaces and can be completely predicted. The dynamics governed by the
time-dependent Schr\"{o}dinger equation is thus reduced to the topology of
the solutions of the time-independent Schr\"{o}dinger equation.

\section{Topology of chirping}

The essence of the adiabatic passage induced by chirping is captured with
the effective two-state Hamiltonian in the rotating wave approximation \cite
{Allen,Shore} 
\begin{equation}
{\sf H}(t)=\frac{\hbar }{2}\left[ 
\begin{array}{cc}
0 & \Omega (t) \\ 
\Omega (t) & 2\Delta (t)
\end{array}
\right] ,
\end{equation}
which describes the radiative interaction between a two-level system (states 
$\left| 1\right\rangle $ and $\left| 2\right\rangle )$ and the
quasi-resonant laser field through the effective Rabi frequency $\Omega (t)$
and the effective detuning $\Delta (t)$. We have assumed that spontaneous
emissions are negligibly small on the time scale of the pulse duration. The
population resides initially in the state $\left| 1\right\rangle $.

In these processes, $\Omega (t)$ stands for a one- or multi-photon Rabi
frequency (depending on the process studied, see e.g. Ref. \cite
{Chelkowski_JRS} for an effective two-photon chirping) and 
\begin{equation}
\Delta (t)=\Delta _{0}(t)+S(t)
\end{equation}
is the sum of the detuning from the one- or multi-photon resonance and of
the effective dynamical Stark shift. This effective dynamical Stark shift $%
S(t)$ results from the difference of the dynamical Stark shifts associated
to the two energy levels and produced by the laser fields non resonant with
the other levels of the system. For the {\it direct chirping}, the detuning
from the resonance $\Delta _{0}(t)$ is time dependent due to an active
sweeping of the laser frequency. The dynamical Stark shifts $S(t)$ are in
general detrimental since they shift the levels away from the resonance. For
the {\it Stark chirping}, the quasi-resonant laser frequency is not chirped
(the detuning $\Delta _{0}$ is time independent); the time dependence of the
effective detuning $\Delta (t)=\Delta _{0}+S(t)$ is only due to Stark shifts
which are induced by the laser pulses \cite{Yatsenko_PRA99,Rickes}.

\begin{figure}[tbp]
\includegraphics{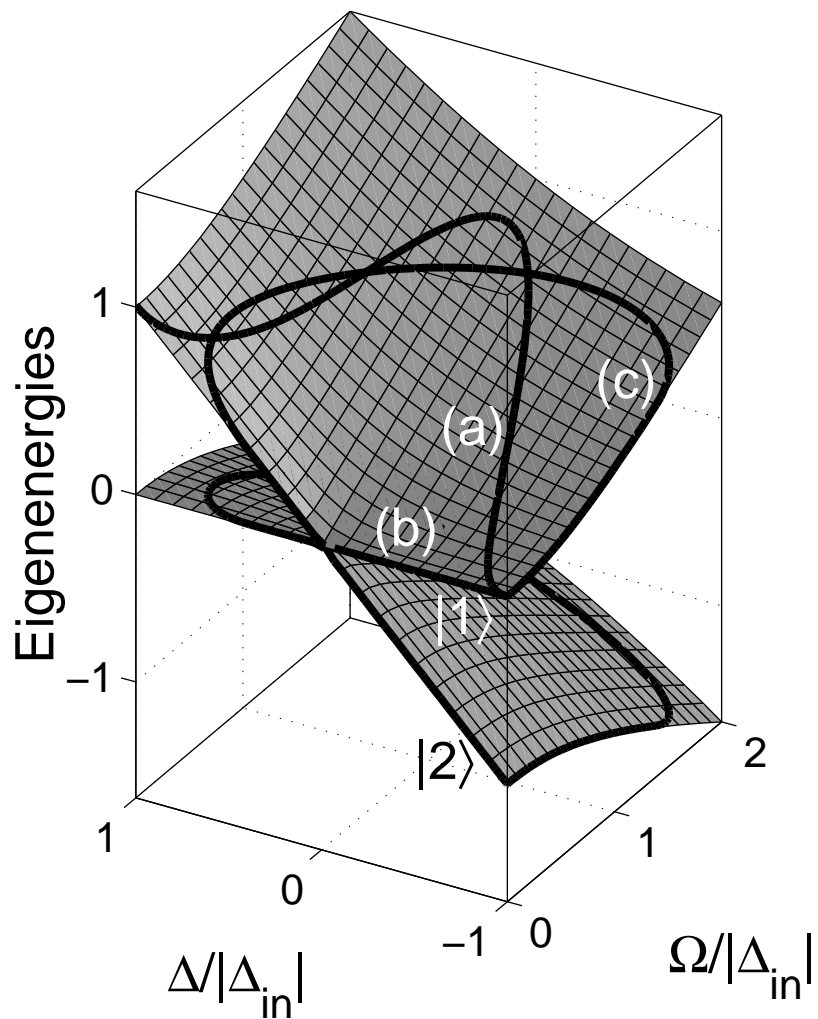}
\caption{Surfaces of eigenenergies (in units of $\left\vert \Delta_{\text{in}%
}\right\vert$) as functions of $\Omega/\left\vert \Delta_{\text{in}%
}\right\vert$ and $\Delta/\left\vert\Delta_{\text{in}}\right\vert$. Three
different paths, denoted (a), (b) and (c) are depicted: (a) corresponds to a
direct chirping and (b) and (c) to SCRAP.}
\label{chirping}
\end{figure}

The process can be completely described by the diagram of the two surfaces 
\begin{equation}
\lambda _{\pm }(\Omega ,\Delta )=\frac{\hbar }{2}\left( \Delta \pm \sqrt{%
\Omega ^{2}+\Delta ^{2}}\right)
\end{equation}
which represent the eigenenergies as functions of the instantaneous
effective Rabi frequency $\Omega $ and detuning $\Delta $ (see Fig. \ref
{chirping}). All the quantities are normalized with respect to a
characteristic detuning denoted $\Delta _{\text{in}}$. They display a
conical intersection for $\Omega =0,\Delta =0$ induced by the crossing of
the lines characterizing the states $\left| 1\right\rangle $ and $\left|
2\right\rangle $ for $\Omega =0$ and various $\Delta $. The way of passing
around or through this conical intersection is the key of the successful
transfer. Three generic curves representing all the possible passages with a
negative initial detuning $-\left| \Delta _{\text{in}}\right| $ are shown.
Note that the three other equivalent curves with a positive initial detuning
have not been drawn. The path (a) corresponds to a direct chirping of the
laser frequency from the initial detuning $-\left| \Delta _{\text{in}%
}\right| $ to the final one $+\left| \Delta _{\text{in}}\right| $. The paths
(b) and (c) correspond to SCRAP with $\Delta _{0}=-\left| \Delta _{\text{in}%
}\right| $. For the path (b), while the quasi-resonant pump pulse is off,
another laser pulse (the Stark pulse, which is far from any resonance in the
system) is switched on and induces positive Stark shifts $S(t)>0$ (the Stark
pulse frequency is chosen with this aim). The Stark pulse makes thus the
eigenstates get closer and induces a resonance with the pump frequency .
This resonance is mute since the pump pulse is still off, which results in
the true crossing in the diagram. The pump pulse is switched on after the
crossing. Later the Stark pulse decreases while the pump pulse is still on.
Finally the pump pulse is switched off. As shown in the diagram, the
adiabatic following of the path (b) induces the complete population transfer
from state $\left| 1\right\rangle $ to state $\left| 2\right\rangle $. The
path (c) leads exactly to the same effect: The pump pulse is switched on
first (making the eigenstates repel each other as shown in the diagram)
before the Stark pulse $S(t)>0$ which is switched off after the pump pulse.

The conditions for adiabatic passage involving one unique statevector are
standard: adiabatic evolution is satisfied when the rate of changes $\left| 
\stackrel{\cdot }{\Theta }(t)\right| $ in the mixing angle $\Theta (t)$,
defined as $\tan 2\Theta (t)=\Omega (t)/\Delta (t),\ -\pi \leq 2\Theta
(t)\leq 0,$ is much smaller than the separation of the eigenvalues $\left|
\lambda _{+}(t)-\lambda _{-}(t)\right| /\hbar =\sqrt{\Omega ^{2}(t)+\Delta
^{2}(t)}:$%
\begin{equation}
\left| \stackrel{\cdot }{\Theta }(t)\right| \ll \sqrt{\Omega ^{2}(t)+\Delta
^{2}(t)}.  \label{adiab}
\end{equation}
The peak amplitudes, the delay between the two fields and the pulse shapes
are chosen such that these conditions of adiabatic passage are met. Detailed
conditions of adiabatic passage can be found in \cite{Rickes} for delayed
Gaussian pulses.

\section{Topology of stimulated Raman adiabatic passage}

\begin{figure}[tbp]
\includegraphics{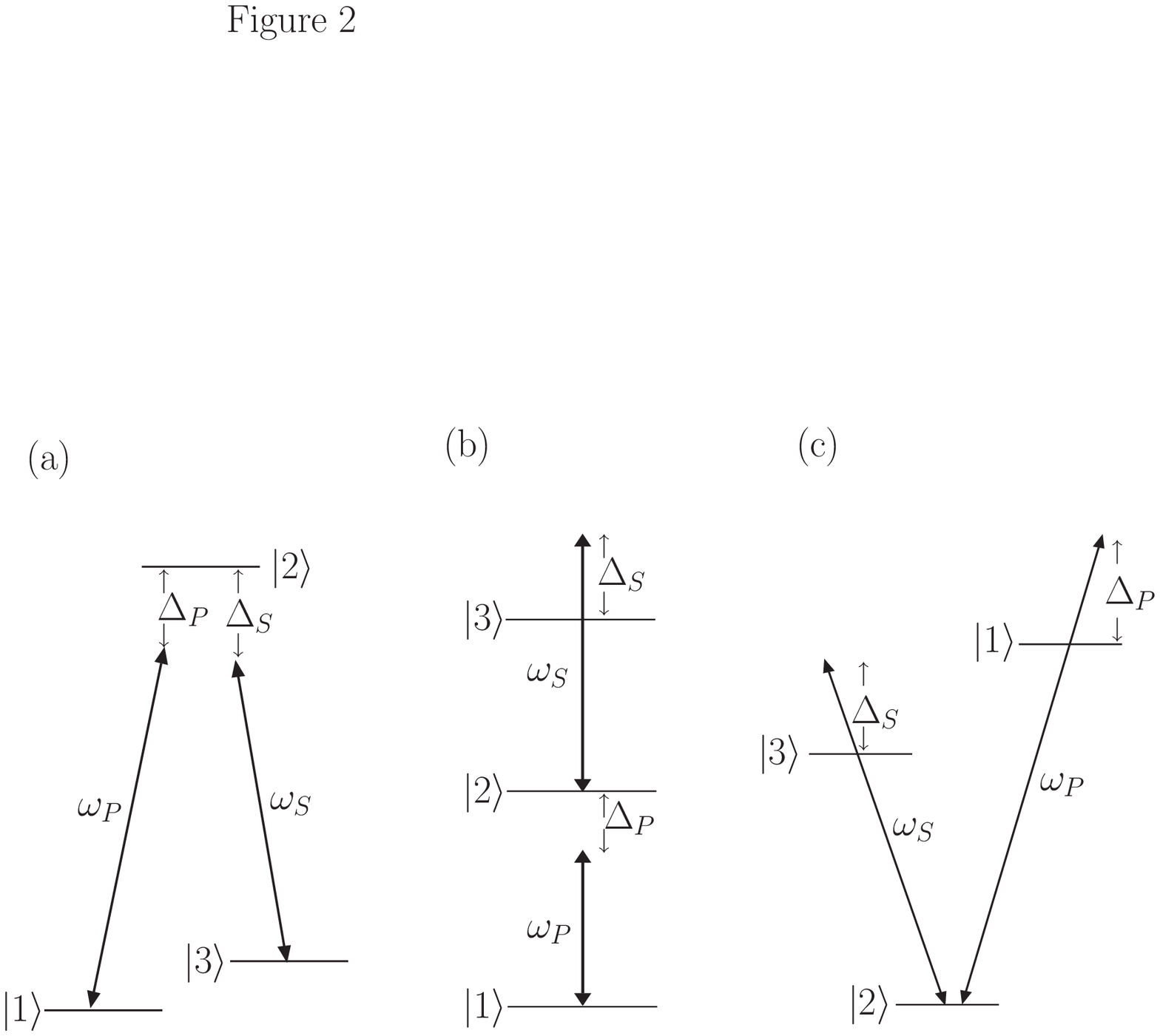}
\caption{Diagram of linkage patterns between three atomic states showing
pump $(P)$ and Stokes $(S)$ transitions and the various detunings for (a) $%
\Lambda$ and (b) $V$ systems.}
\label{linkage}
\end{figure}

The adiabatic passage induced by two delayed laser pulses, the well known
process of STIRAP, produces population transfer in $\Lambda $ systems (see
Fig. \ref{linkage}a). (The pump field couples the transition 1-2 and the
Stokes field couples the transition 2-3.) It is known that, the initial
population being in state $\left| 1\right\rangle $, the complete population
transfer is achieved with delayed pulses, either (i) with a so-called
counterintuitive temporal sequence (Stokes before pump) for various
detunings as identified in Refs \cite{Danileiko_94,STIRAP21_2}, or (ii) with
a two-photon resonant (or quasi-resonant) pulses but far from the one-photon
resonance with the intermediate state $\left| 2\right\rangle $, for any
pulse sequence (demonstrated in the approximation of adiabatic elimination
of the intermediate state \cite{Grischkowsky}). Here we revisit the STIRAP
process through the topology of the associated surfaces of eigenenergies as
functions of the two field amplitudes.

Our results are also valid for ladder and $V$ systems.

We also show the following results which are new to our knowledge: (i) we
can transfer the population to state $\left| 3\right\rangle $ with intuitive
(as with counterintuitive) specific quasi-resonant pulses {\it without
invoking the approximation of adiabatic elimination}, (ii) with specific
quasi-resonant pulses, we can {\it selectively} transfer the population to
state $\left| 2\right\rangle $ for an {\it intuitive} sequence or to state $%
\left| 3\right\rangle $ for a {\it counterintuitive} sequence, and (iii)
with an intuitive or counterintuitive sequence, we can {\it selectively}
transfer the population to state $\left| 2\right\rangle $ or to state $%
\left| 3\right\rangle $ playing on the {\it detunings} and on the{\it \ peak
pulse amplitudes ratio}. We remark that the selectivity (ii) has been
demonstrated in the case of exact two-photon resonant pulses \cite{Ter}.
This last result is however not robust since it depends on using precisely
determined total pulse areas.

We also analyze the counterpart of the previous processes in $V$ systems
(see Fig. \ref{linkage}b): the initial population being in state $\left|
2\right\rangle $, we show that with specific non-resonant pulses, (i) we can 
{\it selectively} transfer the population to state $\left| 1\right\rangle $
for an intuitive sequence or to state $\left| 3\right\rangle $ for a
counterintuitive sequence; (ii) we can {\it selectively} transfer the
population to state $\left| 1\right\rangle $ or to state $\left|
3\right\rangle $ playing on the ratio of the peak pulse amplitudes.

The most general Hamiltonian in the rotating wave approximation for these
processes reads 
\begin{equation}
{\sf H}(t)=\frac{\hbar }{2}\left[ 
\begin{array}{ccc}
0 & \Omega _{P}(t) & 0 \\ 
\Omega _{P}(t) & 2\Delta _{P} & \Omega _{S}(t) \\ 
0 & \Omega _{S}(t) & 2\left( \Delta _{P}-\Delta _{S}\right)
\end{array}
\right] ,
\end{equation}
with $\Omega _{j}(t),\ j=P,S$ the one photon Rabi frequencies associated
respectively to the pump pulse (of carrier frequency $\omega _{P}$) and the
Stokes pulse (of carrier frequency $\omega _{S})$. We have assumed that the
states $\left| 1\right\rangle $ and $\left| 3\right\rangle $ have no dipole
coupling and that spontaneous emission from the upper state $|2\rangle $ is
negligibly small on the time scale of the pulse duration. The rotating wave
transformation is valid if $\Omega _{P}(t)\ll \left| E_{2}-E_{1}\right| $
and $\Omega _{S}(t)\ll \left| E_{3}-E_{2}\right| $, where $E_{j}$, $j=1,2,3$
are the energies associated to the bare states $\left| j\right\rangle $

The detunings $\Delta _{P}$ and $\Delta _{S}$ are one-photon detunings with
respect to the pump and Stokes frequencies respectively and 
\begin{equation}
\delta =\Delta _{P}-\Delta _{S}
\end{equation}
is the two-photon detuning.

\begin{figure}[tbp]
\includegraphics{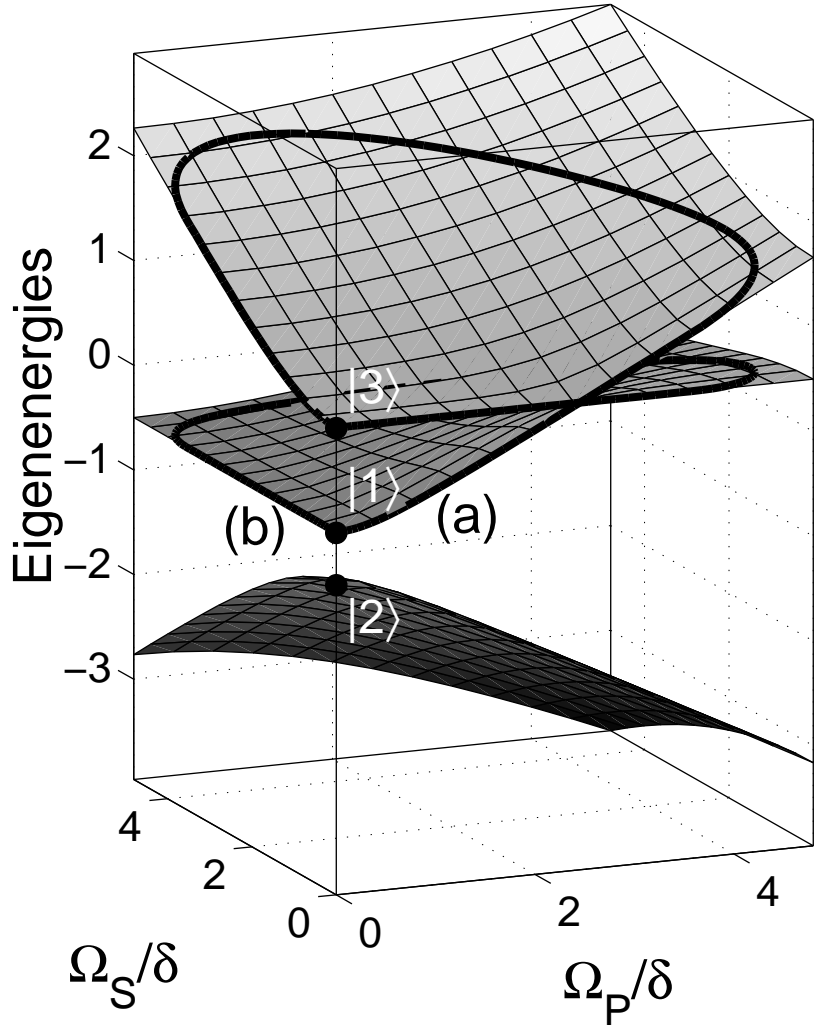}
\caption{Surfaces of eigenenergies (in units of $\protect\delta$) as
functions of $\Omega_P/\protect\delta$ and $\Omega_S/\protect\delta$ for the
case 213. The paths (a) and (b) (constructed with delayed pulses of the same
length and peak amplitude) correspond respectively to the intuitive and
counterintuitive pulse sequences in $\Lambda$ or ladder systems (for which
the initial population resides in state $\vert 1\rangle$).}
\label{F213}
\end{figure}

\begin{figure}[tbp]
\includegraphics{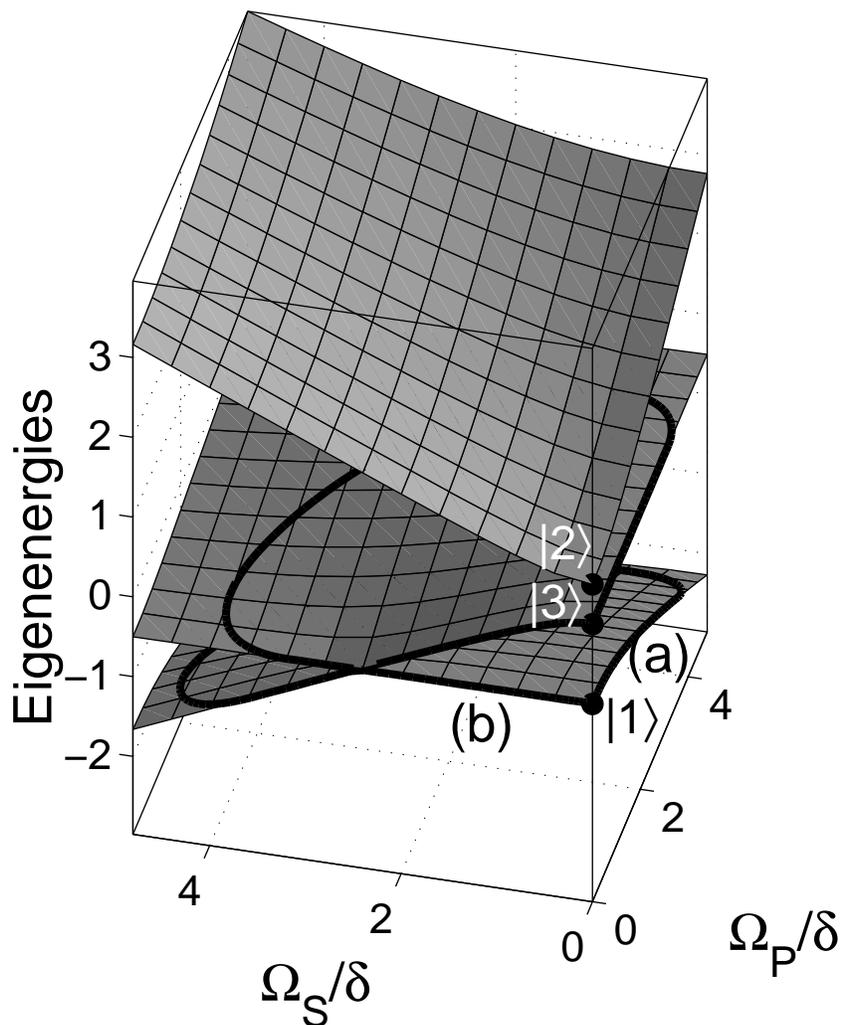}
\caption{Surfaces of eigenenergies (in units of $\protect\delta$) as
functions of $\Omega_P/\protect\delta$ and $\Omega_S/\protect\delta$ for the
case 132. The paths (a) and (b) (with pulses of the same length and peak
amplitude) correspond respectively to the intuitive and counterintuitive
pulse sequences in $\Lambda$ or ladder systems.}
\label{F132}
\end{figure}

For $\Lambda $, ladder and $V$ systems (see respectively Fig. \ref{linkage}%
a, b and c), 
the one-photon detunings $\Delta _{P}$, $\Delta _{S}$ are respectively
defined as 
\begin{mathletters}
\begin{eqnarray}
\hbar \Delta _{P} &=&E_{2}-E_{1}-\hbar \omega _{P},\quad \hbar \Delta
_{S}=E_{2}-E_{3}-\hbar \omega _{S}, \\
\hbar \Delta _{P} &=&E_{2}-E_{1}-\hbar \omega _{P},\quad \hbar \Delta
_{S}=E_{2}-E_{3}+\hbar \omega _{S}, \\
\hbar \Delta _{P} &=&E_{2}-E_{1}+\hbar \omega _{P},\quad \hbar \Delta
_{S}=E_{2}-E_{3}+\hbar \omega _{S}.
\end{eqnarray}
In what follows we study the topology of the eigenenergy surfaces for
various generic sets of the parameters. The topology depends on the
detunings which determine the relative position of the energies at the
origin. We study various {\it quasi-resonant} pulses in the sense that the
detunings are small with respect to the associated peak Rabi frequencies,
i.e. 
\end{mathletters}
\begin{mathletters}
\begin{eqnarray}
\Delta _{P} &\lesssim &\max_{t}(\Omega _{P}),\quad \Delta _{S}\lesssim
\max_{t}(\Omega _{S}), \\
\delta &\lesssim &\max_{t}(\Omega _{P}),\quad \quad \delta \lesssim
\max_{t}(\Omega _{S}).
\end{eqnarray}

Allowing large enough amplitudes imply three generic cases for $\delta >0$,
which are referred to 213, 132 and 123 (these number sets are associated to
the eigenenergies for zero field amplitudes from the smallest to the
biggest). Three other symmetric and thus equivalent cases (referred as 312
symmetric with 213, 231 with 132 and 321 with 123) appear for $\delta <0.$

\subsection{The cases 213 and 132}

The case 213 corresponds to $\Delta _{S}<\Delta _{P}<0$ and its symmetric
312 to $0<\Delta _{P}<\Delta _{S}.$ One example for the case 213 is
diagrammed in Fig. \ref{F213} for $\Delta _{P}=-\delta /2$ and $\Delta
_{S}=-3\delta /2$.

The case 132 corresponds to $0<\Delta _{S}<\Delta _{P}$ (and its symmetric
231 to $\Delta _{P}<\Delta _{S}<0$) as diagrammed in Fig. \ref{F132} for $%
\Delta _{P}=3\delta /2$ and $\Delta _{S}=\delta /2$. Fig. \ref{F132} shows
that the topology of the 132 case is similar to the topology of the 213
case. In both cases, the surface continuously connected to the state $%
|2\rangle $ is isolated from the two other surfaces which present a conical
intersection for $\Omega _{S}=0$ (resp. $\Omega _{P}=0$) in the 213
configuration (resp. 132 configuration). This crossing corresponds to a mute
resonance as described above for chirping. The topologies shown on the
respective figures \ref{F213} and \ref{F132} are generic for the condition 
\end{mathletters}
\begin{equation}
\Delta _{P}\Delta _{S}>0,  \label{213}
\end{equation}
with respectively 
\begin{equation}
\left| \Delta _{P}\right| <\left| \Delta _{S}\right| \quad \text{and}\quad
\left| \Delta _{P}\right| >\left| \Delta _{S}\right| .
\end{equation}

In the following, we describe in detail the 213 case (see Fig. \ref{F213}).
For the process in $\Lambda $ or ladder systems, where the initial
population resides in state $|1\rangle $, two different adiabatic paths lead
to the complete population transfer, depending on the pulse sequence. The
path denoted (a) corresponds to an intuitive sequence for the increasing
pulses. The pump pulse is switched on first, making the levels connected to
the states $|1\rangle $ and $|2\rangle $ repel each other (dynamical Stark
shift) until the level connected to $|1\rangle $ crosses the level connected
to $|3\rangle $. The Stokes pulse is switched on after the crossing. Next
the two pulses can decrease in any sequence. The path (b) is associated to a
counterintuitive sequence for the decreasing pulses. The two pulses can be
switched on for any sequence. The pump pulse has to decrease through the
crossing when the Stokes pulse is already off. These two results are valid
even without application of adiabatic elimination. The conditions of
adiabaticity are very similar to the ones of the chirping case (\ref{adiab}).

The $V$ systems are uninteresting in these cases since the final population
comes back to the state $|2\rangle $ for any pulse sequence.

\subsection{The case 123}

\begin{figure}[tbp]
\includegraphics{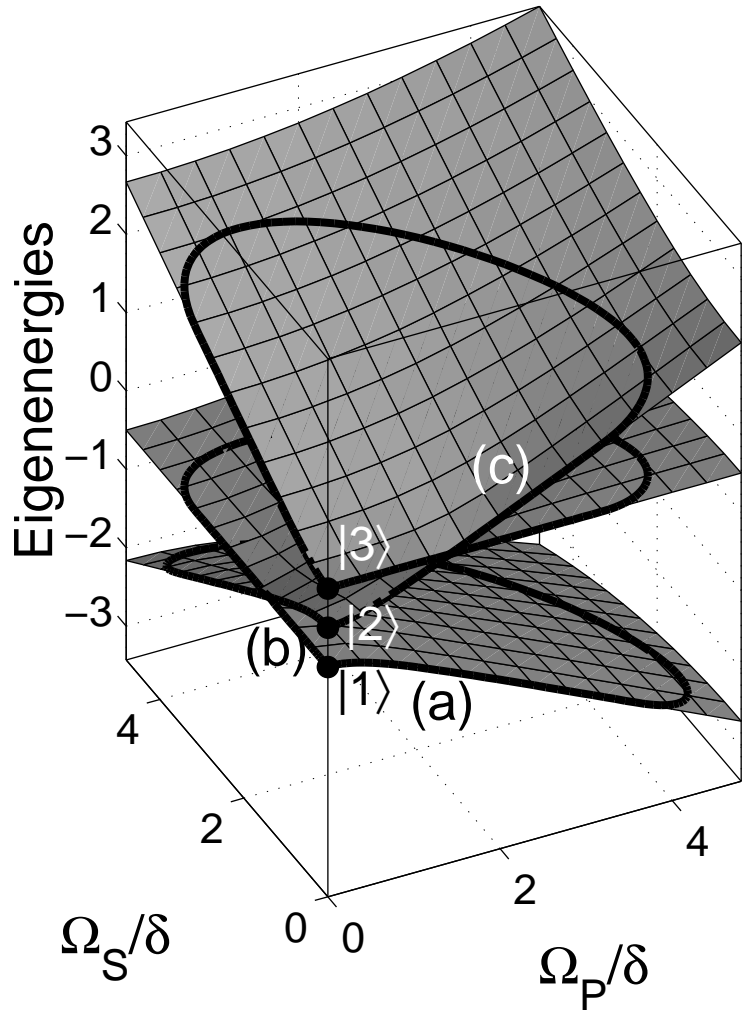}
\caption{Surfaces of eigenenergies (in units of $\protect\delta$) as
functions of $\Omega_P/\protect\delta$ and $\Omega_S/\protect\delta$ for the
case 123. The paths (a) and (b) (with pulses of the same length and peak
amplitude) correspond respectively to the intuitive (transfer to $\vert
2\rangle$) and counterintuitive (transfer to $\vert 3\rangle$) pulse
sequences in $\Lambda$ or ladder systems leading to the selective transfer.
The paths (a) and (c) correspond to the selective transfer in $V$ systems
(for which the initial population resides in $\vert 2\rangle$), respectively
to $\vert 1\rangle$ and $\vert 3\rangle$.}
\label{F123}
\end{figure}

The case 123 corresponds to $\Delta _{S}<0<\Delta _{P}$ and its symmetric
321 to $\Delta _{P}<0<\Delta _{S}.$ One example for the case 123 is
diagrammed in Figs \ref{F123} for $\Delta _{P}=\delta /2$ and $\Delta
_{S}=-\delta /2$. The topology shown on this figure is generic for the
condition 
\begin{equation}
\Delta _{P}\Delta _{S}<0.  \label{123}
\end{equation}
In this configuration, two conical intersections involve the intermediate
surface, one with the lower surface and another with the upper surface. This
topology gives here more possibilities for transfer: {\it the combined
choice of the pulse sequence and the ratio of the peak amplitudes allows the
selective transfer into the two other states}.

Figure \ref{F123} shows that, for the process in $\Lambda $ (or ladder)
systems, two different adiabatic paths lead to different complete population
transfers, depending on the pulse sequence. The path (a) characterizes an
intuitive pulse sequence (for decreasing pulses) and allows to populate at
the end the state $|2\rangle $. The Stokes and pump pulses are switched on
in any sequence and the pump pulse is switched off before the Stokes. The
paths (b) characterizes a counterintuitive pulse sequence (for increasing
pulses) and allows to populate at the end the state $|3\rangle $. The Stokes
pulse is switched on before the pump and the pulses are switched off in any
sequence. We can thus selectively populate the states $|2\rangle $ or $%
|3\rangle $ provided the peak amplitudes are sufficiently strong to induce
the adiabatic path to cross the intersection involved.

For the process in $V$ systems, the paths (a) and (c) of Fig. \ref{F123}
show the respective selective transfer into the states $|1\rangle $ or $%
|3\rangle $.

\begin{figure}[tbp]
\includegraphics{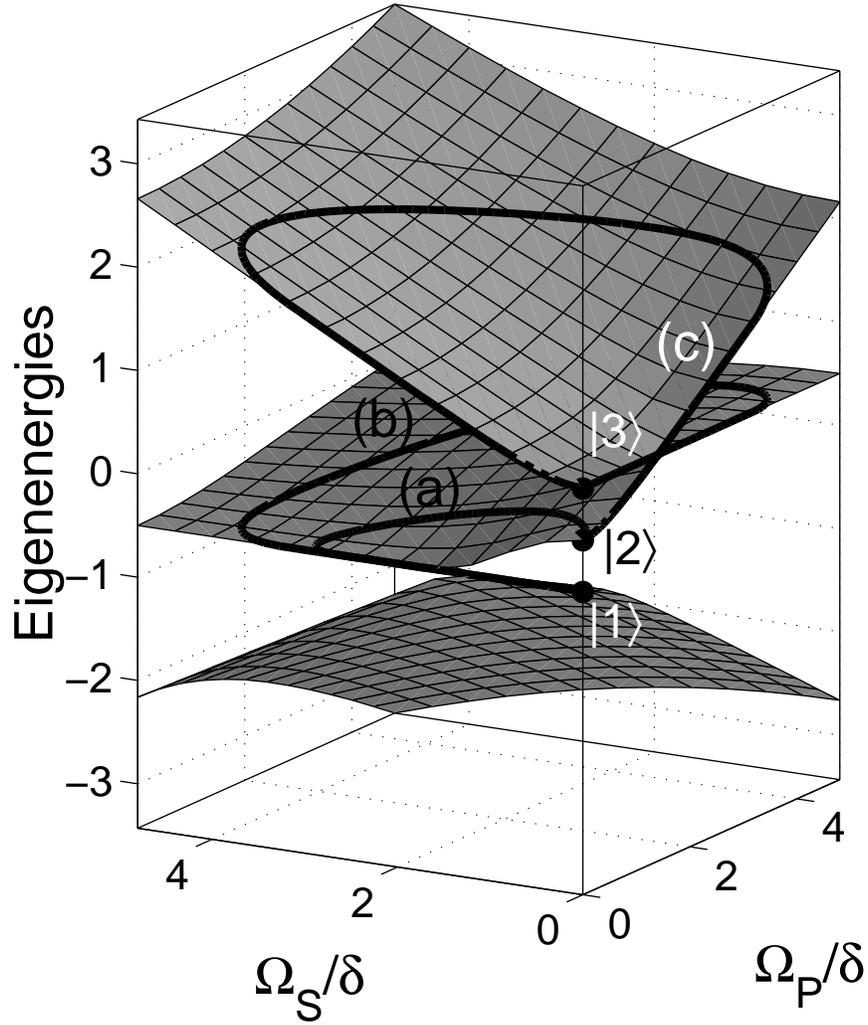}
\caption{Surfaces of eigenenergies with the same parameters as Fig. \ref
{F123} showing the selective transfer with pulses of different peak
amplitudes and length. For counterintuitive sequences in $\Lambda$ or ladder
systems, the path (b) [corresponding to the path (b) of Fig. \ref{F123}]
shows the transfer to $\vert 3\rangle$, and the path (a) (with pulses of
different length and peak amplitude) characterizes the transfer to $\vert
2\rangle$. The paths (a) and (c) correspond to the selective transfer in $V$
systems.}
\label{selectivity}
\end{figure}

Figure \ref{selectivity} corresponds to the same topology of Fig. \ref{F123}
but with a different path (a). Figure \ref{selectivity} shows that, for $%
\Lambda $ (or ladder) systems with counterintuitive sequences, we can
selectively populate the states $|2\rangle $ or $|3\rangle $ if the pulse
sequence are designed differently in their sequence and their peak
amplitude. The paths (b) corresponds to the previous path (b) of Fig. \ref
{F123} and allows to populate at the end the state $|3\rangle $. The path
(a) is characterized by a pump pulse (still switched on after the Stokes
pulse) longer and of smaller peak amplitude and allows to populate at the
end the state $|2\rangle $. Note that we can obtain a similar path (a) with
a counterintuitive pulse sequence and equal peak amplitudes if the detuning $%
\Delta _{P}$ is taken smaller so that the crossing for $\Omega _{S}=0$ is
pushed to higher pump pulse amplitude $\Omega _{P}$.

For $V$ systems, Fig. \ref{selectivity} shows that this selectivity [paths
(a) and (c)] also occurs (for any sequence of the pulse).

\section{Discussion and conclusions}

In this article we have applied simple geometrical tools to two- and
three-level systems in the rotating wave approximation to classify all the
possibilities of complete population transfer by adiabatic passage, when the
two-level system is driven by a chirped laser pulse and the three-level
system by two delayed pulses. We have shown that the complete transfer by
adiabatic passage is intrinsically related to the topology of the
eigenenergy surfaces. We have found the following new results in the
three-level systems such as, in $\Lambda $ or ladder systems, (i) robust
population transfer to the state $|3\rangle $ by an intuitive sequence of
quasi-resonant pulses, (ii) robust selective transfer to the states $%
|2\rangle $ and $|3\rangle $ depending on the design of the pulses (lengths,
amplitudes and delay).

\begin{figure}[tbp]
\includegraphics{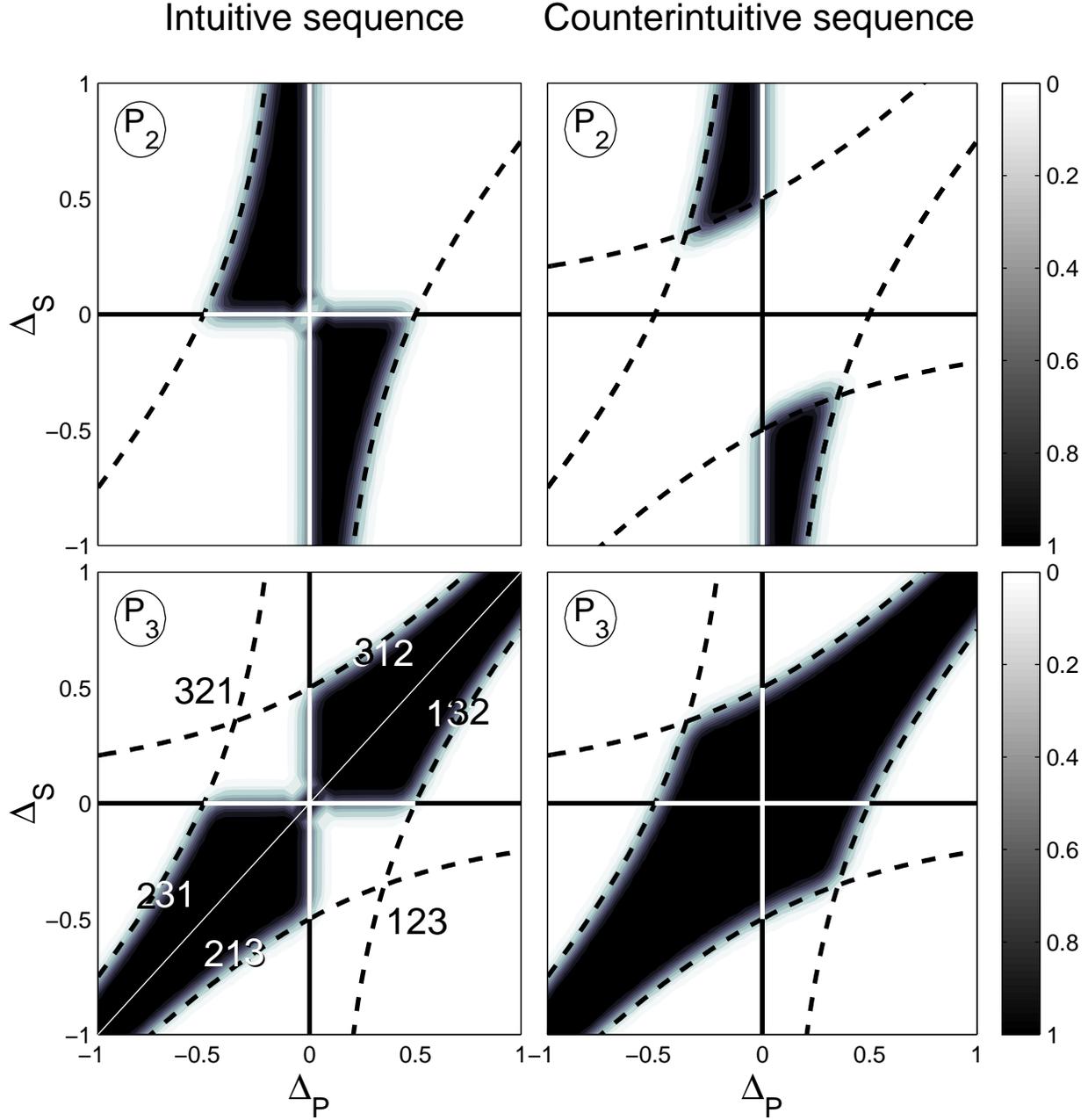}
\caption{Transfer efficiencies $P_2$ to $\vert 2\rangle$ (upper row) and $%
P_3 $ to $\vert 3\rangle$ (lower row) at the end of the pulses for the
intuitive (left column) and counterintuitive (right column) sequences of
delayed sine-squared pulses with the same peak amplitude $\Omega _{\max }$
and a large temporal area $\Omega _{\max }\protect\tau =500$ ($\protect\tau$
is the pulse length and the delay is $\protect\tau /2$). The efficient
population transfers are bounded by $\Delta_P=0$ and $\Delta_S=0$ (thick
full lines) and the branches of hyperbolas (dashed lines). The areas bounded
by the full lines are labelled by the cases 213, 132, 123, $\ldots$ The
three first ones correspond respectively to Figs \ref{F213}, \ref{F132} and 
\ref{F123}.}
\label{simul}
\end{figure}

The topology gives information on the dynamics for purely adiabatic passage.
For real pulses of finite duration one has to complement these information
with the analysis of the effects of non-adiabatic corrections. Figure \ref
{simul} shows numerical calculations that illustrate some of the predictions
of the analysis of section III. Its displays the populations of the states $%
|2\rangle $ and $|3\rangle $ at the end of the pulses for intuitive and
counterintuitive sequences with a large pulse area. The boundaries of the
areas of efficient transfer (black areas) are predicted quite accurately by
the topology analysis: They are determined by (i) the straight lines (thick
full lines) $\Delta _{P}=0$ and $\Delta _{S}=0$ coming from the inequalities
(\ref{213}) and (\ref{123}) and (ii) the branches of the hyperbolas (dashed
lines) 
\begin{eqnarray}
\Delta _{S} &=&\Delta _{P}-\frac{\left( \Omega _{\max }\right) ^{2}}{4\Delta
_{P}}, \\
\Delta _{P} &=&\Delta _{S}-\frac{\left( \Omega _{\max }\right) ^{2}}{4\Delta
_{S}},
\end{eqnarray}
which are determined from the positions of the conical intersections. Figure 
\ref{simul} shows that the efficiency of the robust population transfer to
the states $|2\rangle $ or $|3\rangle $ is identical for the intuitive and
counterintuitive sequences except in two regions: (i) areas bounded by $%
\Delta _{P}\Delta _{S}<0$ and the branches of the hyperbolas, where the
population is transferred in a robust way to state $|2\rangle $ for the
intuitive sequence or to state $|3\rangle $ for the counterintuitive
sequence and (ii) a area (smaller for longer pulse areas) near the origin
where non adiabatic effects are strong for the intuitive sequence and where
the population transfer depends precisely on the pulse areas for this
intuitive sequence (see the comments below). Non adiabatic effects, which
are smaller for larger pulse areas, also occur near the boundary regions.

For the concrete realization with finite pulses of moderate areas, we have
to analyze the precise influence of non adiabatic effects. In the following
we study these non adiabatic effects referring to Fig. \ref{F213} supposing
that the detunings are small enough with respect to the speed of the process
to yield non adiabatic transitions.

In the intuitive case, at the beginning of the process, the states $%
|1\rangle $ and $|2\rangle $ are coupled by the pump pulse, and thus non
adiabatic transitions can occur near the origin between the surfaces
connected to $|1\rangle $ and $|2\rangle $. In the counterintuitive case, at
the beginning of the process, state $|1\rangle $ is not coupled to the other
levels and there are no non adiabatic transitions near the origin. At the
end of the process, the adiabatic path ending in $|3\rangle $ is not coupled
to the other levels, implying again absence of non adiabatic transitions
near the origin. We thus recover the well known fact that resonant STIRAP is
more favorable with counterintuitive pulse sequence and leads to Rabi
oscillations in the intuitive case.

The consequences of the topology on the population transfer with exact
resonances at $\Omega _{S}=0$, $\Omega _{P}=0$ giving rise to degeneracies
will be discussed in a forthcoming work.

\section{Acknowledgments}

We acknowledge support by INTAS 99-00019. LY thanks l'Universit\'{e} de
Bourgogne for the invitation during which this work was accomplished. We
thank K. Bergmann and R. G. Unanyan for useful discussions.

%

\end{document}